\documentclass[12pt,preprint]{aastex}
\shorttitle{Secondary bar of LMC}
\shortauthors{A. Subramaniam}
\begin{document}
\title{Evidence of a mis-aligned secondary bar in the Large Magellanic Cloud}
\author{Annapurni Subramaniam\altaffilmark{1}}
\affil{Indian Institute of Astrophysics, Koramangala II Block, Bangalore - 34}
\begin{abstract}
Evidence of a mis-aligned secondary bar, within the primary bar of the
Large Magellanic Cloud (LMC) is presented.  The density distribution and the
de-reddened mean magnitudes ($I_0$) of the
red clump stars in the bar obtained from the OGLE II data are used for this study.
The bar region which predominantly showed wavy pattern
in the line of sight in \citet{a03} was located. These points in the X-Z plane 
delineate an S-shaped pattern, clearly indicating a mis-aligned bar. This feature is
statistically significant and does not depend on the considered value of $I_0$ for the LMC center.
The rest of the bar region were not found to show the warp or the wavy pattern.
The secondary bar is found to be considerably elongated in the Z-direction, 
with an inclination of 66$^o$.5 $\pm$ 0$^o$.9, whereas the undisturbed part of the
primary bar is found to have an inclination of 15$^o$.1 $\pm$ 2$^o$.7, such that
the eastern sides are closer to us with respect to the western sides of both the bars. 
The PA$_{maj}$  of the secondary bar is found to be
108$^o$.4 $\pm$ 7$^o$.3. 
The streaming motions found in the H I velocity map close to the LMC center could be
caused by the secondary bar. The recent star formation and the gas distribution in LMC
could be driven by the mis-aligned secondary bar.
\end{abstract}
\keywords{galaxies: Magellanic Clouds -- galaxies: stellar content, structure}
\section{Introduction}
The off-centered stellar bar is one of the most striking features of the 
Large Magellanic Cloud (LMC). On the other hand, this is one of the least studied and
understood feature of the LMC. The near-IR star count maps presented by \citet{v01} 
found the bar to be a smooth structure,
even though a {\it peak} in the ellipticity and change in position angle (PA) were
found within the central 2$^o$.
Recently \citet{a03} studied the relative distance within the LMC bar using the de-reddened
mean magnitudes of red-clump stars
and found that the bar is warped and also found structures in the bar. In an attempt to find
out the possible reason for these structures, we came across
evidence of a possible existence of a secondary bar within the LMC bar. 
Bars are a common phenomenon in late-type spirals and Magellanic irregulars \citep{df73}. Recent studies
find that the
secondary bars within large-scale bars are also common, occurring in about a third of barred galaxies
\citep{J97, l02, es02, E03}.

The evidence of a possible existence of the secondary bar has been found in the literature.
Some of the significant references are discussed below.
In the R-band isophotes of \citet{d57}, the first contour shows the bar, two contours immediately
next to this suggests a turn in the top-left and bottom-right corners of the bar.
This is the first evidence for the twist of the isophotes within the central region of the LMC bar.
The isophotes are based on R-band photometry and hence contribution from the old stars dominate.
Such an isophotal twist would  manifest as 
change in the PA of the major axis in the central region, change in 
ellipticity and a possible counter-rotation in the inner regions of the LMC. 
The evidence of the change in PA and ellipticity near the LMC center can be found in the literature
and some are indicated below. Figure 3 in
\citet{v01} shows the change in PA and ellipticity, figure 6 in
\citet{vahs02} shows change in PA for the carbon stars, figure 5 in \citet{k98} shows the
change in PA for H I. 
%The change in ellipticity in the central region is found by \citet{v01}.
The presence of the above two features found in the central region of NGC 2950 is 
taken as the photometric signature of a misaligned secondary bar \citep{c03}.
Recent investigations of double-barred galaxies \citep{E03, J97} indicate that, 
in general, the radial
plot of the ellipticity reflects double peak corresponding to both the bars in the galaxy.
In the case of LMC, figure 3 in \citet{v01}, indicates the first peak, $\epsilon_{max} \sim 0.7$
at r$\sim$1.$^o$0. This is well within the primary bar, which extends to more than 2$^o$ radius.
The second peak appears after a radial distance of 2$^o$.0, though it is not very prominent.
The corresponding $\epsilon_{max} \sim 0.57$.
Evidence of negative rotational velocity with respect to the
center of LMC is noticed in a number of cases.
The study of CH stars by \citet{HC88} found that some stars have negative galactocentric 
velocity. Similar cases are also found in the case of planetary nebulae and old star clusters.
\citet{HC88} state that these stars may be related in someway to the bar of the LMC.
Recent studies of carbon star kinematics by \citet{vahs02} find that within the central 1$^o$, 
the mean rotational velocity is $\sim$ $-$28 Km/s. 
Thus all the above observations point to the possible existence of a misaligned secondary bar. 
All these are features observed in the projected two dimensions and no information is available
on its possible appearance in the line of sight.

In the present study, we explore the presence of the secondary bar within the LMC bar in the 
projected two dimensional X-Y plane as well as in the X-Z plane. We use the red clump stars
in the OGLE II catalogue as the probe for this study. The density of 
red clump stars in the bar region is used to study the projected pattern. The relative 
distance estimates in the LMC bar based on the de-reddened mean magnitudes of red clump stars, 
presented in \citet{a03} 
are used to study the pattern in the line of sight.

\section{Bar in the projected two dimensions}
OGLE II survey \citep{u00} consists of photometric data of 7 million stars in B, V and I 
pass bands in the central 5.7 square degree of LMC.
The observed bar region was divided into 1344 sections (3.56$\times$3.56 arcmin$^2$).
The red clump stars were identified using I vs (V$-$I) colour-magnitude
diagram (CMD) and on an average, 2000 red clump stars were identified per region.
The data suffers from the incompleteness problem due to crowding effects, and 
the incompleteness in the data in I and V pass bands are tabulated in \citet{u00}.
After correcting for the data incompleteness, the total number of red clump stars
in each area bin  and the number density of the red clump stars per sq. degree
were estimated.
The center of the LMC is taken to be $05^h19^m38^s.0$ $-69^o27'5".2$ (J2000)
\citep{df73}. The location of each area bin is converted to the linear X, Y coordinates, 
using the convention in \citet{vc01} and this data is used for the following analysis.

\begin{figure}
\plotone{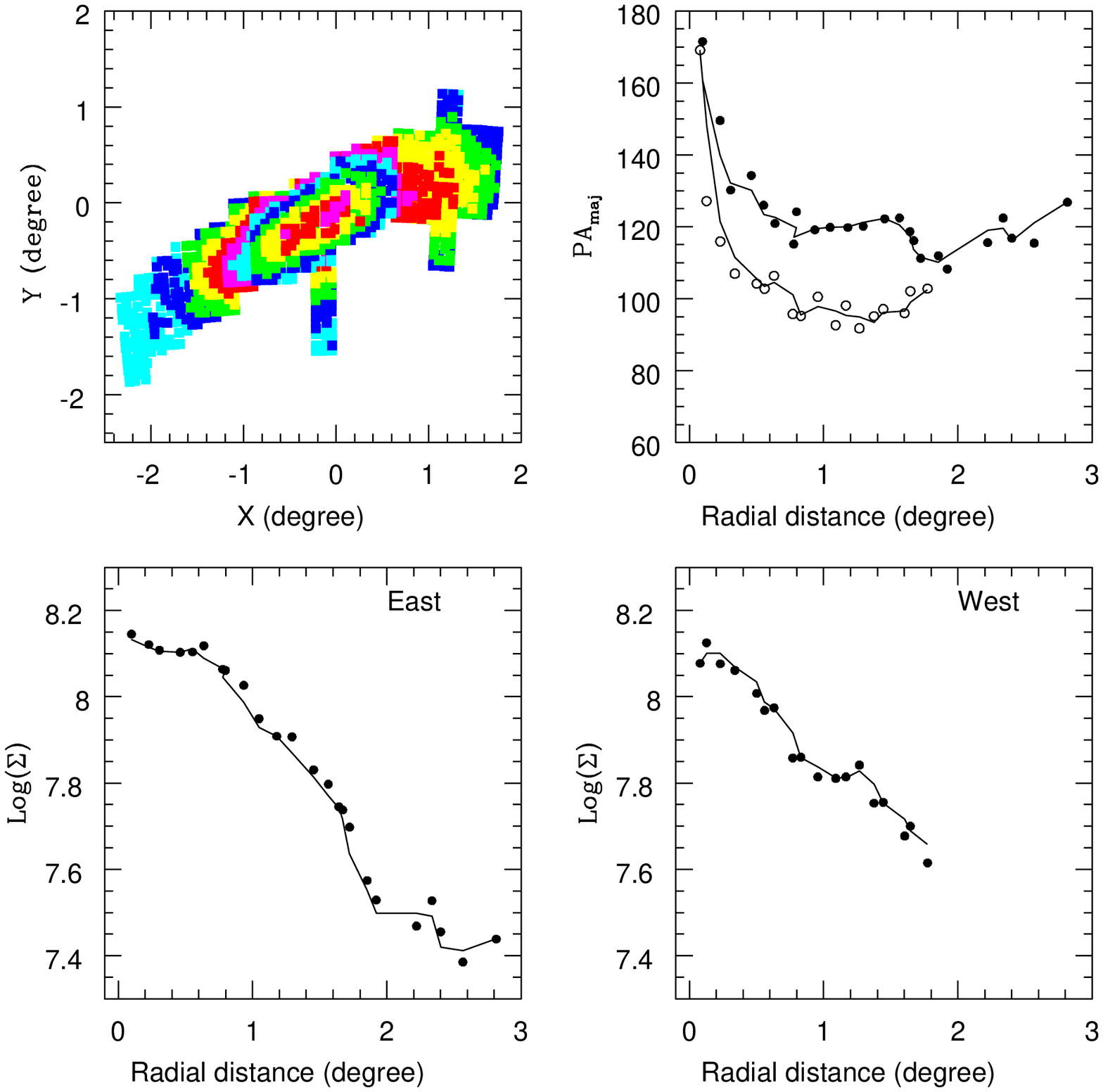}
\caption{ The density distribution of the red clump stars is presented in the top left panel.
 The central region indicated in magenta
has density more than 1.3 X 10$^8$ stars per sq. degree. The colour code used is such that
the density decreases from magenta to cyan, with density intervals of 0.1 X 10$^8$ stars per sq. degree.
The second appearance of magenta has density bet ween 0.8\,10$^8$ and 0.7 X 10$^8$, and the points
indicated in cyan at the east end of the bar has density less than 0.3 X 10$^8$. 
The variation of PA of the major axis with radial distance is shown for the east side (filled circles)
and west side (open circles) of the bar in the top right panel. A smoothing function of width 2 
is used to connect the points.
The radial variation of the maximum value of the red clump density is shown for the
east (left) and west (right) sides of the bar in the bottom panel. The points are connected using
a smoothing function, which takes the average of two adjoining points. 
\label{fig1}}
\end{figure}

The two-dimensional distribution of the red clump star density on the X-Y plane
is shown in the top left panel of figure \ref{fig1}. The figure 
shows maximum density near the center which decreases radially outwards.
The main feature is the elongation of the central density to the eastern side, 
and this elongation is then carried outward as ellipses.
This is found to be the origin of the elliptical pattern found in the bar. The major-axis
of the ellipse is found to turn very clearly in the east side. This is indicative of isophotal twist.
The maximum density at each radial point is estimated and its variation with respect to radius is
shown in the bottom panels of Figure \ref{fig1}. The profiles are different for the east and 
west sides of the bar.
The profile on the east side is characterised by a shallower slope upto 0$^o$.6 and a steeper
slope upto 1$^o$.9. Beyond this, the profile is found to be very flat.
The other features which could also be noticed
are the likely hood of a rise close to the center, at 0$^o$.6 and 1$^o$.3. 
The rise of the profile near the center could indicate a very compact bulge,
which could not be confirmed here. The points are connected using
a smoothing function which takes average of the two neighbouring points. The profile for the east
side is very similar
to the magnitude variation along the major axis 
of a double barred galaxy (eg. NGC 1291 \citet{d75}, figure 10 ). The prominent peak corresponding
to the bulge is missing here with the rest of the profile looking very similar.
NGC 1291 is considered as a prototype of double-barred galaxies \citep{fm93}.
The west side profile is different, with the
slope similar to that seen between 0$^o$.6 and 1$^o$.9 for the eastern side and a rise 
in the profile at 1$^o$.3. The variation of the PA of the major axis as a function 
of radial distance is shown in the top right panel of Figure \ref{fig1}.
Considerable change in the PA with radial distance can be noticed.
Turn over of the PA at a radial distances of 0$^o$.8, 
1$^o$.4 and 1$^o$.9 on the eastern side and 1$^o$.3 on the western side could be noticed.
Thus the turns over of the PA is well correlated with the changes in the density profile.
These variations thus suggest structures in the bar close
to the center, with the possible existence of a secondary bar.
The PA of the maximum density points are used to estimate the average PA$_{maj}$ of the bar.
The average value of the PA of the major axis is found to be 123$^o$.3 $\pm$ 13$^o$.3 for the east
and 105$^o$.5 $\pm$ 18$^o$.1 for the west side. The average PA of the primary bar is found to be
114$^o$.0 $\pm$ 22$^o$.5. This value tallies well with 
the earlier estimates of the PA$_{maj}$ for inner regions, which is dominated by the primary bar.
The value of PA$_{maj}$ is 122$^o$.5 $\pm$ 8$^o$.3  
as estimated by \citet{v01} and 129$^o$.9 $\pm$ 6$^o$.0
\citet{vahs02}.

\section{The location of the bar in the Z-direction}
The results presented in \citet{a03} showed that there are structures in the bar 
indicating a wavy
pattern running from the east to the west side of the bar. This was noticed on and above the
warp, where the western end as well as the eastern end were found to be closer to us with respect
to the LMC center. The data presented in \cite{a03} were transferred to the X-Y plane using the
assumed LMC center. The points at which the bar was found to be located away from us, 
(ie., at RA $\sim 84 ^o$; Dec $\sim - 70^o$ \& LMC center \& RA $\sim 76^o$; Dec $\sim -69^o$) 
are found to lie in one line. The position angle of the axis coinciding with this line is 
109$^o \pm$ 3$^o$. 

\begin{figure}
\plotone{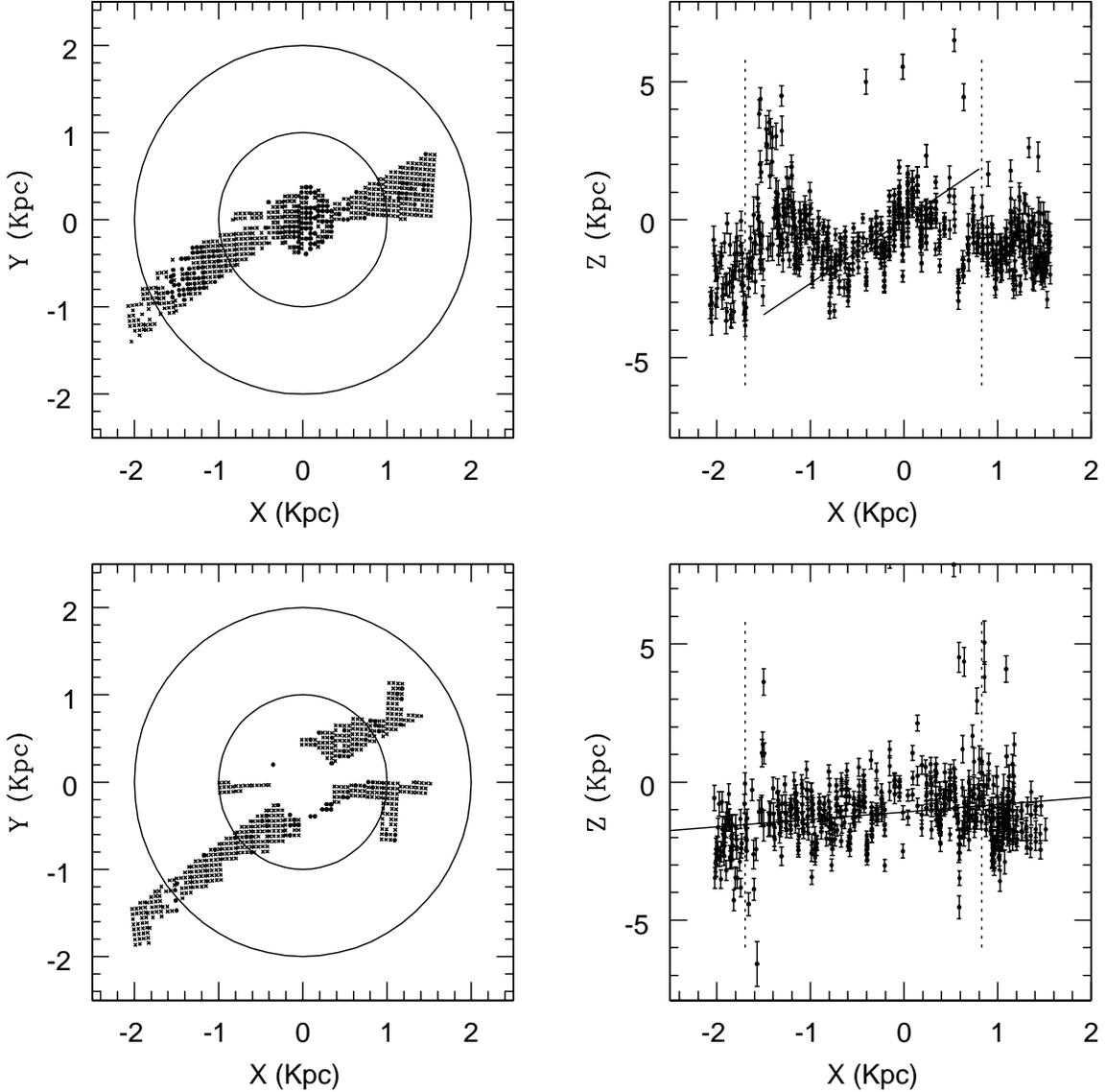}
\caption{The location of the primary bar within the PA = $110^o.0 \pm 12^o.0$ is shown in the
X-Y and X-Z plane in the top panel. The location of the rest of the bar in the X-Y and X-Z
planes are shown in the bottom panel. The circles in the left figures are drawn at 1$^o$.0 and
2$^o$ radii. The filled circles in the left panel indicate points which have Z$>$ 0.0.
The dotted lines shown in the right figures correspond to locations showing
change in density profiles. The errors in the data points are obtained from \citet{a03}, where
the $\Delta I_0$ values are converted to Kpc as explained in the text.
\label{fig2}}
\end{figure}

The figure 3 of \citet{a03} was converted into the X-Z plane, where the Z-values were
obtained by converting the $I_0$ values. The $I_0$ value at the location of 
the LMC center, 18.20 $\pm$0.01 mag was taken as Z=0. The difference in the value of
$I_0$ between any location and the LMC center is estimated and then this value
is multiplied by 25 Kpc, since a shift in 0.1 mag in I corresponds to a shift of 2.5 Kpc.
Since Z is in Kpc, we convert X also in Kpc using the relation, 1 degree = 0.89 Kpc.
The magnitude errors in the data points were also converted to errors in distance.
As the bar is seen to be perturbed along the PA, 109$^o$.0, the locations between $\pm$12$^o$.0
of this PA were chosen. All the points within a radius of 0.$^o$4 were also included as the
above selection results in severe under sampling of data points near the center. 
These points are shown in the top panel of Figure \ref{fig2}, where, the
left panel shows the X-Y plane and the right panel shows the X-Z plane. In the X-Y plane, the
points with Z $> 0$ are shown as filled circles. 
In the X-Z plane,
the dotted lines correspond to the location were a change in radial density profile was
noticed in Figure \ref{fig1}, at radial distances of 1$^o$.9 and 1$^o$.3 in the east and 
west sides respectively. Within these two lines, a well defined S-pattern can be noticed. The variations
seen in the feature is statistically significant, as indicated by the error bars. 
The maximum random error as estimated by \citet{a03} was 0.02 mag in $I_0$ and this corresponds
to 500 pc as per the above conversion. The variation as shown in the figure corresponds to
more than 1.5 Kpc, which has 3$\sigma$ significance with respect to the maximum random error.
The feature found in the X-Z plane is not dependent on the $I_0$ value chosen for the center.
The same pattern is observed when the $I_0$ magnitudes are plotted against X coordinate.
Hence this feature does not depend on the value of $I_0$ considered for the conversion to the
Z-axis. 

The S-pattern consists of the central bar inclined in the line of sight with trailing pattern 
on both ends. This feature is considered as the mis-aligned secondary bar. 
The eastern end of the secondary bar is seen to be closer with respect to the western end. 
Thus a part of the primary bar is disturbed due to the presence
of the secondary bar. 
On both ends of the secondary bar, primary bar is sheared and has a depth 
of about 5 Kpc. The nature of rotation of the bar could not be inferred from the present data,
though the S-pattern gives an impression of rotation in the 
counter-clockwise direction. The central feature corresponding to the bar is fitted with a 
straight line as shown in the figure. The inclination of the bar is estimated from the
slope and is found to be 66$^o$.5 $\pm$ 0$^o$.9. The value of the slope depends on the points
chosen and the above value was derived by obtaining the best value of correlation coefficient (0.78)
for the least square fit. A deviation of $\pm$ 5$^o$.0 for the slope is possible based on the
choice of points. The PA of the major axis of the
secondary bar is also estimated. The value was found to have a lot of scatter for points near the
center. We estimated PA$_{maj}$ to be 108$^o$.4 $\pm$ 7$^o$.3, for the bar outside 
0$^o$.4 radius and 136$^o$.2 $\pm$ 26$^o$.0 for points within. The PA of the
secondary is computed by taking only those data points which are used to fit the straight line.
The PA of the secondary bar can be considered to be 108$^o$.4 $\pm$ 7$^o$.3, where the contribution
from the central regions is not considered.
The extent of the bar could not be estimated accurately as the trailing patterns overlap.
Also, the spiral pattern is clearly seen to be connected
to the bar in the eastern side, but a clear connection is not seen in the western side.
The upper limit to the length of the bar is estimated to be $\sim$ 3.0 Kpc.
As indicated above, only a part of the bar is chosen to identify the
secondary bar. This would mean that the rest of the bar should not show any variation in the
line of sight. The lower panel of figure \ref{fig2} shows the X-Y and X-Z plot of rest of the bar. 
It is very clear from the X-Z plot that the rest of the bar does not show any wavy pattern 
or the warp. The data points are fitted with a straight line and is shown in the figure. 
The unperturbed part of the primary bar is thus slightly inclined,
with a slope of 0.27 $\pm$0.05, which corresponds to 15$^o$.1$\pm 2^o$.7. Thus the east
end is closer to us than the west end of the primary bar.

\section{Results and Discussion}
The possible existence of a secondary bar within the primary bar of LMC is explored here.
The photometric signatures of the secondary bar is found in plentiful in the literature, like
the twist of isophotes, change in the PA of the major axis and ellipticity {\it peak} in the central regions.
On the other hand, these signatures were never connected with the possible existence of a secondary bar.
The motivation to look for a secondary bar came from the perturbations that were noticed
in the primary bar by \citet{a03}. The radial profile of the maximum density on the east side,
resembles the brightness profile of the double barred galaxies along the major axis. 
The secondary bar is seen only on the east side. This indicates that the 
secondary bar is not symmetric with respect to the optical center.
The secondary bar has disturbed only a part of the primary bar, hence we are also able to estimate the
parameters of the undisturbed primary bar. The undisturbed primary bar does not show any east
west asymmetry. 
Though the ellipticity of the bars could not be estimated here, the ellipticity estimations 
in the literature shows that the ellipticity of the secondary bar is $\epsilon^s_{max} \sim 0.7$,
whereas that of the primary is $\epsilon^p_{max} \sim 0.57$. The catalog of double-barred galaxies 
presented by \citet{E03} indicates that the average value of the ellipticity of the secondary bar
in 49 galaxies is 0.3, whereas the average value for the primary bar is 0.47. The above values
were found to be similar to the average ellipticity of the bars presented in \citet{J97} for
13 galaxies. Both the data also indicate that $\sim$ 85\% galaxies show higher ellipticity 
value for the primary bar when compared to that of the secondary bar. On the other hand, LMC shows
higher ellipticity for the secondary bar, which is seen in $\sim$ 15\% of the double-barred galaxies.
It can be seen that the difference between the PAs of the primary and the secondary bar is very small.
The $\Delta PA = 8^o.0 \pm 23^o$.0, which is very small, or close to zero within errors. LMC
belongs to the group of 6\% of double-barred galaxies, which show very small value for 
$\Delta PA$ \citep{E03}.
The bars are not aligned in the Z-direction, as indicated by the inclination values
of 66$^o$.5 $\pm$ 0$^o$.9 and 15$^o$.1 $\pm$ 2$^o$.7 for the primary and the 
secondary bars respectively. The main signature which reveals the central structure as a secondary
bar is the mis-alignment in the Z-direction, more than the ellipticity and isophotal signatures.
This is the reason why the central structure is claimed to be a 
 mis-aligned secondary bar. The spiral like patterns on the ends of the secondary bar could 
suggest a possible counter-clockwise rotation in the X-Z plane. 
The presence of a mis-aligned secondary bar could give rise to kinematic signatures
near the central regions. The negative rotational velocity noticed in the central regions 
could be due to the secondary bar. The mis-alignment could also produce non-circular motions.
As the feature in the X-Z plane does not show any ring, either the
secondary bar has a slow pattern speed or it is recently formed. 
More studies are required to understand this newly found feature in the LMC.

The signatures of the
secondary bar could be traced in H I observations.
The velocity field of the H I, as shown in figure 4, of \citet{k98}, indicates a steep
velocity gradient just to the north of the center of the bar. This is considered as the
dynamical evidence of large scale streaming motions. Such steep velocity gradient
was also noticed by \citet{lr92}. It is quite possible that the secondary bar
is responsible for the streaming motions. The elongation of the secondary bar in the
Z-direction could give rise to a steep velocity gradient.
The H I observations by \citet{k98} show a two armed spiral pattern in their figure 2.
Similar spiral arm features were also observed by \citet{ss03},
figure 2, and they remark that the two arms are connected by H I
but not in a structure that looks like the bar as the position of the optical bar is different.
The secondary bar could be the feature which is connecting the two spiral arms. The H I
observation of \citet{r84} find clear indications of non-circular velocity near the center.
In figure 5 of \citet{r84} and figure 8 of \citet{lr92}, the velocity map of H I clearly
indicates a lower velocity to the south of the thickly populated iso-velocity contours
near the LMC center, and a higher velocity to the north. This suggests a rotation in the sense
that the southern part is moving towards us and the northern part is moving away with respect to
the center. This corresponds to counter-clockwise rotation in the X-Z plane. 
If the secondary bar has a counter-clockwise rotation, then this is in 
good agreement with the rotation seen in the H I velocity maps.
Thus the secondary bar of the LMC could be the missing link between the stellar and the gas
distribution in the LMC. The recent star formation and the gas distribution in the LMC
could be driven by this mis-aligned secondary bar.

I thank T.P.Prabhu and Daniela Vergani for helpful discussions.

\clearpage
\clearpage

\clearpage

\end{document}